\documentclass[a4paper,10pt]{article}
\usepackage[a4paper,hmargin=15mm,top=15mm]{geometry}
\usepackage{textcomp} 
\usepackage{amsmath} 
\usepackage{pstool} 
\usepackage{multicol} 
\usepackage[colorlinks=true]{hyperref} 
 \makeatletter

\newenvironment{figurehere}
  {\def\@captype{figure}}
  {}
\makeatother

\title{Multi-octave supercontinuum from mid-IR filamentation in bulk}
\author{F. Silva$^{1,*}$, D. Austin$^{1}$, A. Thai$^{1}$, M. Baudisch$^{1}$, M. Hemmer$^{1}$, A. Couairon$^{2}$ and J.~Biegert$^{1,3}$\\
$^1$ICFO-Institut de Ciences Fotoniques, Mediterranean Technology Park\\\
 08860 Castelldefels (Barcelona), Spain\\
$^2$Centre de Physique Th\'{e}orique, CNRS, \'{E}cole Polytechnique, F-91128 Palaiseau, France \\
$^3$ICREA - Instituci\'{o} Catalana de Recerca i Estudis Avan\c{c}ats, 08010 Barcelona, Spain
}
\date{}

\begin{document}

\maketitle

\begin{multicols}{2}

{\bf 
In supercontinuum generation, various propagation effects combine to produce a dramatic spectral broadening~\cite{Alfano-2006-supercontinuum} of intense ultrashort optical pulses with far reaching possibilities~\cite{Jones2000}. Different applications place highly divergent and challenging demands on source characteristics such as spectral coverage from the ultraviolet (UV)~\cite{Akozbek-2006-Extending,Ernsting-2001-Wave} across the visible (VIS) to the near-infrared (NIR), and into the mid-infrared (MIR)~\cite{Domachuk-2008-Over,Qin-2009-Ultrabroadband,Kasparian-2000-Infrared,Theberge-2008-Ultrabroadband}. Shot-to-shot repeatability, high spectral energy density, an absence of complicated or non-deterministic pulse splitting are also essential for many applications. Here we present an ``all in one'' solution with the first supercontinuum in bulk covering the broadest bandwidth from just above UV far into the MIR. The spectrum spans more than three octaves, carries high spectral energy density (3\,pJ up to 10\,nJ per nanometer), has high shot-to-shot reproducibility, and is carrier-to-envelope phase (CEP) stable. Our method, based on filamentation~\cite{Braun95} of a femtosecond MIR pulse in the anomalous dispersion regime, allows for a new class of simple and compact supercontinuum sources.
}

Coherent ultra-broadband light sources are in demand for a large variety of applications. Modern imaging --- especially optical coherence tomography~\cite{Huang1991} --- exploits broadband light sources to map biological materials with high spatial resolution.
Multi-photon spectroscopy and microscopy~\cite{Denk1990a} demand focusable and coherent light at a variety of wavelengths to visualize cells or for chemical imaging of cellular components. 
Modern nonlinear spectroscopy \cite{Mukamel-1995-Principles} benefits from time-resolved multi-color pump-probe capabilities for tracking  the motion of valence electrons and nuclei on their natural femtosecond timescale.
Breath analysis is a promising technique for early cancer detection demanding spectroscopic detection of volatiles which is achieved with high sensitivity in the molecular fingerprint regions in the MIR~\cite{Murtz-2008-Online}.
Frequency combs~\cite{Udem1998,Jones2000} in the VIS to MIR enable ultrahigh resolution spectroscopy used, for example, to study the composition of extrasolar planets. Finally, when combined with advances in pulse shaping, coherent supercontinua enable arbitrary multi-color waveform synthesis of spectra spanning the UV to MIR, which offers the possibility of simultaneously probing and controlling~\cite{Brumer2003} electronic, vibrational and rotational motion, or synthesizing single electric field waveforms at arbitrary carrier wavelengths~\cite{Krauss-2010-Synthesis}.

These applications place widely varying demands on the characteristics of a supercontinuum (SC) source. 
Tremendous progress has been made using novel optical fibers, with SC spanning the UV into the MIR~\cite{Domachuk-2008-Over,Qin-2009-Ultrabroadband,Xia-2006-Mid}. However, besides the intrinsic peak power limitations and alignment sensitivity, the most extreme bandwidths observed in fibers have been achieved at the expense of pulse to pulse repeatability, which often manifests in complicated temporal pulse splitting translating into spectral modulation~\cite{Ranka-1996-Observation,Qin-2009-Ultrabroadband}. A requirement for shot-to-shot repeatability and a smooth spectrum therefore limits the allowable propagation length, restricting the bandwidth significantly~\cite{Qin-2009-Ultrabroadband}. 

A maturing understanding of the mechanisms of filamentation \cite{Couairon-2007-Femtosecond} reveals the difficulty of achieving significant spectral broadening to the long wavelength side of the pump: most of the underlying processes tend to blue shift the spectrum. One contribution to the spectral broadening is due to the free electrons which are produced by ionization throughout the pulse. These produce an ultrafast decrease in the refractive index, leading to a blue shift. The effect may be clamped by ionization losses, dispersion, and defocusing. Another mechanism for spectral broadening is self-phase modulation, which produces a spectral shift proportional to the temporal intensity gradient, with red shift on the rising edge and blue on a falling edge. A key role, therefore, is played by temporal reshaping of the pulse. Two of the main sources of temporal reshaping --- self-steepening and space-time focusing --- intrinsically produce a steep trailing edge, again resulting in a blue shift. Other contributions to reshaping --- ionization-induced defocusing and loss --- also tend to act more strongly on the trailing edge and centre of the pulse respectively.
Steep intensity gradients or shocks are naturally arrested by dispersion.
Which of the mechanisms dominate depends strongly on the pulse and material parameters~\cite{Moll-2004-Role, Brodeur-1998-Band,Gaeta-2000-Catastrophic}.
Despite its pessimistic outlook for achieving significant spectral broadening to longer wavelengths, the modern elucidation of bulk material filamentation does raise the tantalizing prospect that, since the clamping of the blue shift is largely determined by the material properties, the short wavelength edge of the spectrum could be largely independent of the pump wavelength, offering the potential for enormous spectra when pumping in the MIR. More rigorous theoretical support for this conjecture comes from a Born approximation-like view of SC in bulk materials \cite{Kolesik-2008-Perturbative} in which the spectral extent is largely determined by the medium dispersion. In this view, the angularly-resolved ($\{\theta,\lambda\}$) spectrum is linked to the mechanism through phase-matching of ``x-waves''. Normal, near-zero or anomalous dispersion manifests in ``x'', ``fish'', or ``o'' shaped structures~\cite{Couairon2006,Hernandez2008} in the $\{\theta,\lambda\}$ domain.
{\begin{figure*}[] 
   \begin{center}
  \includegraphics{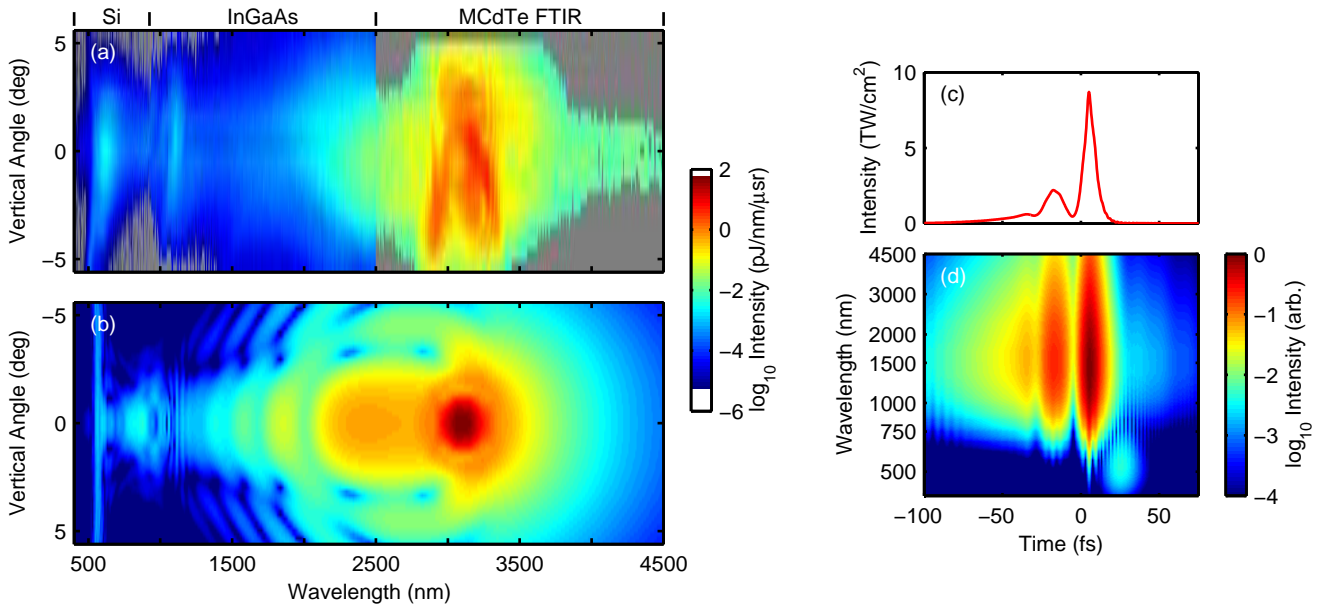} 
   \caption{ \label{fig:2d} (a) Measurement and (b) simulation of the far field $\{\theta,\lambda\}$ representation in absolute units; the grey area in (a) indicates the noise level and the detector ranges are shown above the figure. (c) Temporal intensity profile and (d) spectrogram of the on-axis simulated pulse emerging from the YAG plate; note the logarithmic intensity colour and wavelength scale in the spectrogram.}
   \end{center}
\end{figure*}}

Here we present the broadest supercontinuum ever produced in bulk, from filamentation of a femtosecond MIR pulse in a thin plate of yttrium aluminium garnet (YAG). 
We used our homebuilt optical parametric chirped pulse amplifier (OPCPA), which delivered CEP stable, 10\,$\mu$J energy, 85\,fs pulses at 3100\,nm at 160\,kHz repetition rate~\cite{Thai2011}. 6.9\,$\mu$J of the output was collimated to 6.8\,mm and focused by a 75\,mm focal length CaF$_2$ lens to a spot size of 50 $\mu$m. The peak power was 76 MW which is three times above the critical power for YAG. Beam sizes are given as radius at the $1/e^{2}$ value of the peak intensity. The 2\,mm thick YAG plate was placed into the focal plane (i.e. the front surface is located 1\,mm from the focal plane), producing a filament inside the YAG plate which is clearly visible to the naked eye. We measured a $\{\theta,\lambda\}$ spectrum, the main challenge being the tremendously large wavelength range. We covered the spectrum by simultaneously using three intensity-calibrated fiber-coupled spectrometers. They consisted of a silicon CCD spectrometer for wavelengths up to 925\,nm, an InGaAs CCD spectrometer for 925--2500 nm, and a Fourier-transform infrared spectrometer (FTIR) with MCdTe detector for wavelengths above 2500\,nm. The three fibers were held parallel to each other with their tips separated horizontally by roughly a millimeter. They were then scanned vertically across the beam, with spectra being acquired at each position. An image registration step in the processing accounted for the different positions of the fibre tips. 	

Figure~\ref{fig:2d}(a) shows a measured $\{\theta,\lambda\}$ spectrum. The long wavelength edge decays smoothly, reaching the noise level  at 4500\,nm, whilst the short wavelength edge has a sharp peak at 610\,nm before dropping rapidly into the noise at 400\,nm. Several structures are present which we analyze below. To obtain higher sensitivity and also demonstrate that all spectral components of the SC may be practically collected and relayed, we refocused the SC into each spectrometer fibre with a 50\,mm lens. The resulting spatially averaged spectrum is shown in Fig.~\ref{fig:1d}. This measurement featured a lower pulse energy but a tighter focus, leading to a slightly different spectrum, and we used different detectors above 2500\,nm as indicated.

The highest spectral energy density is found around the MIR pump wavelength with a maximum energy content of 10\,nJ/nm. The smallest energy content is observed in the 750--1000\,nm range at a few pJ/nm; note that these energies are still sufficient for most absorption spectroscopy experiments. 
Figure~\ref{fig:2d}(a) clearly shows a ``fish-tail'' structure at 610\,nm in the $\{\theta,\lambda\}$ representation whose energy content is roughly one order of magnitude higher than at the minimum at 850 nm. A comparison with theory shows that this feature is not related to odd harmonics of the MIR driver; rather it is consistent with the aforementioned dynamics of x-waves. A small peak can also be observed at around 1100\,nm, which simulations show contains a strong contribution from the third harmonic of the pump.

\begin{figurehere}
\centering
   \includegraphics[]{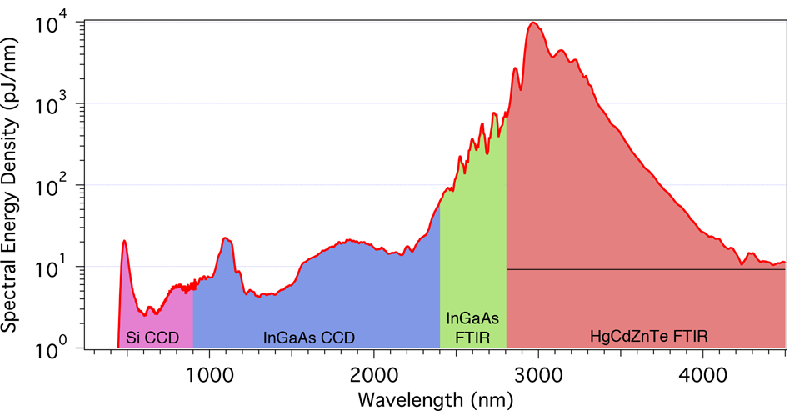}
   \caption{\label{fig:1d}Supercontinuum generated by  3100\,nm, 2.6\,$\mu$J pulses in YAG. 
   The ranges of each spectrometer/detector are indicated, as well as the noise level for the long wavelength detector (grey horizontal line).}
 \end{figurehere}

A practical issue for applications of a SC is the shot-to-shot reproducibility and coherence across the spectrum. The process must therefore be deterministic and not excessively sensitive to the inevitable shot-to-shot fluctuations of the drive laser. These properties have been demonstrated in bulk supercontinuum generation with 800\,nm pumping~\cite{Baltuska-2003-Phase-controlled,Bellini-2000-Phase,Baum-2003-Phase}, and we now show that they also hold for our MIR pumping conditions using an $f$--$2f$ interferometer. In the interferometer, the filament was refocused into a type I beta barium borate (BBO) crystal  resulting in second harmonic generation from a wavelength range of $\approx (1500\pm300)$\,nm. 
The frequency-doubled component interfered with the short-wavelength edge of the SC at $\approx (750\pm150)$\,nm, producing spectral fringes acquired with a spectrometer. Figure~\ref{fig:CEP} shows a 15\,s subset of such a dataset recorded over a 10\,minute interval. The amplitude of the observed fluctuations provides an upper bound on both of the potential sources of phase jitter.  Over the full 10\,minutes we obtain a standard deviation of 259\,mrad. 
\begin{figurehere}
\begin{center}
\includegraphics{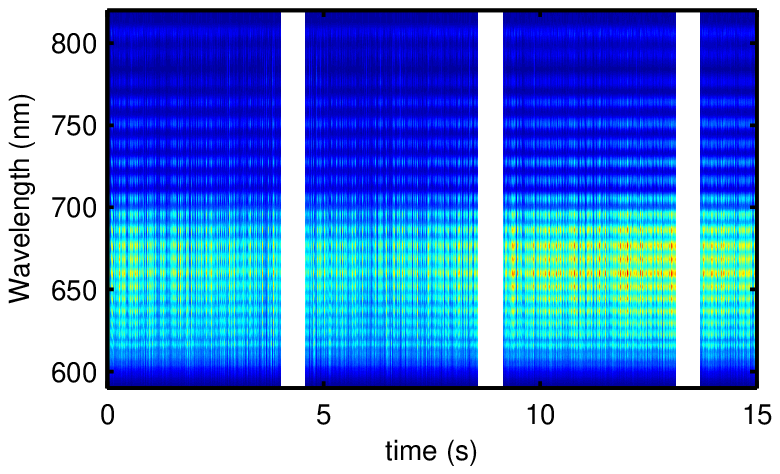}
\caption{\label{fig:CEP}Spectral interference fringes produced by supercontinuum in $f-2f$ interferometer versus time.}
\end{center}
\end{figurehere}

Figure~\ref{fig:2d}(b) shows the $\{\theta, \lambda\}$ spectrum obtained from numerical simulations of the nonlinear propagation of the IR laser pulse in the 2\,mm long YAG plate. A comparison with Fig.~\ref{fig:2d}(a) shows very good qualitative agreement between measurements and simulations. The most intense part of the spectrum corresponds to the ``o''-like structure typical of the anomalous dispersion regime. The hyperbolic branch crossing the axis around 560--620\,nm corresponds to the tail of the ``fish''-wave, i.e., the loci of phase matched angle and wavelength lying in the visible region of the spectrum. Simulations show that this tail appears in the last 400\,$\mu$m of the YAG plate and is not affected by third harmonic generation (THG). Switching off THG in simulations by e.g. propagating the envelope rather than the carrier resolved field only suppress the peak emerging around 1000\,nm. This also indicates that CEP does not affect the ``fish'' wave structure of the spectrum. These features are always present and start with a slightly different focal position with respect to the entrance face of the YAG plate. They also appear with different Kerr coefficients and ionization rates, the accuracy of which is not known with great certainty. Increasing $n_2$ up to a factor of 4 leads to a broader extent of the main spectral peak centered at 3000\,nm and a change in the balance between THG and the SC. Higher Kerr coefficients lead to a prevailing ``fish'' tail with respect to the third harmonic. 

The good agreement between experiment and simulation permits a deeper study of the temporal dynamics using the latter. Figure~\ref{fig:2d}(c) and (d) show the emerging on-axis temporal pulse profiles. Figure~\ref{fig:2d}(c) shows that pulse splitting has not occurred over the nonlinear propagation distance since the main pulse has developed a {pre-pulse} which is not split off completely from the main pulse. These two parts are separated by only a few femtoseconds and preceded by a shallow increasing background. A more detailed picture is revealed in the spectrogram (Fig.~\ref{fig:2d}(d)) with a logarithmic intensity color scaling. We find that the short wavelength extension --- ending in the ``fish tail'' --- consists of a low-intensity trailing edge at 25\,fs. The very weak spectral modulation observed in the simulated $\{\theta, \lambda\}$ representation between 1000--2000\,nm is attributed to the interference of the main pulse with its trailing pedestal. The amplitude of this pedestal is however critically dependent on propagation length and uncertain nonlinear parameters such as the nonlinear refractive index and the ionization rate. We clearly do not observe these weak modulations in our measurement; we attribute this to an absent or much less pronounced pre-pulse or pedestal.

In this Letter we demonstrated the first stable multi-octave SC from filamentation of MIR femtosecond pulses in bulk material. We measured a smooth spectrum from 450\,nm to 4500\,nm, corresponding to 3.3 octaves, with a spectral energy density of 2\,pJ/nm to 10\,nJ/nm. The smoothness of the spectrum indicates absence of chaotic and multiple pulse splitting and a CEP measurement indicated good shot-to-shot coherence over a significant spectral range. We obtained an angularly-resolved far-field spectrum which shows good agreement with full 3D simulations. These indicate that the mechanism is dominated for formation of an ``o'' wave, with the short wavelength edge described by phase-matching. By demonstrating a simple and robust method for coherently extending the spectrum of an amplified femtosecond MIR pulse to the UV we provide a powerful platform for the next generation of ultrafast applications.

{\small
\renewcommand{\refname}{\normalsize References}
\bibliographystyle{brief-with-titles-5authors}
\bibliography{DArefs,jbrefs}

\section*{Acknowledgements}
We acknowledge partial support from the Spanish Ministry of Education and Science through its Consolider Program Science (SAUUL CSD 2007-00013), through Plan Nacional (FIS2008-06368-C02-01), and the Catalan Agncia de Gesti— dÔAjuts Universitaris i de Recerca (AGAUR) with SGR 2009-2013. This research has been partially supported by Fundaci— Cellex Barcelona, and funding from LASERLAB-EUROPE, grant agreement 228334, is gratefully acknowledged. D.A. was supported by Marie Curie Intra-European Fellowship 276556-BAXHHG.

\section*{Author contributions}

F.S. carried out the experiments, with the assistance of A.T., M.B. and M.H.. The data analysis was mainly carried out by D.A. and the simulations were performed by A.C. The experiment was supervised by J.B. and the experimental results were interpreted and written up by D.A. and J.B.
}

\end{multicols}
\end{document}